\documentclass[prd,aps,10pt,twocolumn,floats,nofootinbib,eqsecnum,showpacs]{revtex4-1}

\usepackage{amssymb,amsmath}
\usepackage{epsfig}
\pdfoutput=1
\usepackage{amssymb,amsmath}
\usepackage{epsfig}
\usepackage{xcolor}
\usepackage[utf8]{inputenc}
\usepackage[normalem]{ulem}

\newcommand{\be}{\begin{equation}}
\newcommand{\ee}{\end{equation}}
\newcommand{\ba}{\begin{eqnarray}}
\newcommand{\ea}{\end{eqnarray}}
\newcommand{\beg}{\begin{gather*}}
\newcommand{\eng}{\end{gather*}}
\newcommand{\hh}{,\hspace{0.5cm}}
\newcommand{\hhh}{,\hspace{0.2cm}}

\newcommand{\n}[1]{\label{#1}}

\newcommand{\CAL}{\mathcal}

\usepackage[makeroom]{cancel}
\usepackage[caption=false]{subfig}
\usepackage[colorlinks=true,
            citecolor=green,
            linkcolor=red,
            filecolor=cyan,
            urlcolor=magenta,
            backref=false]{hyperref}

\begin{document}

\title{Classical mechanics with inequality constraints}

\author{Andrei Frolov}
\email{frolov@sfu.ca}
\affiliation{Simon Fraser University, Department of Physics,
Burnaby BC Canada
V5A 1S6 }

\author{Valeri P. Frolov}
\email{vfrolov@ualberta.ca}
\affiliation{Theoretical Physics Institute, University of Alberta, Edmonton, Alberta, Canada T6G 2E1}


\begin{abstract}
In this paper we discuss mechanical systems with inequality constraints. We demonstrate how such constraints can be taken into account by proper modification of the action which describes the  original unconstrained  dynamics. To illustrate this approach we consider a harmonic oscillator in the model with limiting velocity. We compare the behavior of such an oscillator with the behavior of a relativistic oscillator and demonstrated that when an amplitude of the oscillator is large the properties of both type of oscillators are quite similar . We also briefly discuss inequality constraints which contain higher derivatives.
\end{abstract}


\maketitle


\section{Introduction}

Studying  mechanical systems with constrains has a long history (see for example  \cite{goldstein2002,arnold89}). For such systems besides the Lagrangian $L(q,\dot{q})$ there exist one or more constraint relations $\Phi_i=0$ which restrict a motion of the system.   In the simplest case when these functions depend only on the coordinates $q$  the constraints are called holonomic. By solving these constraints one can reduce the configuration space and obtain a reduces Lagrangian which describes the motion of the system in the presence of the constraints. A case of nonholonomic constraints, where constraint functions $\Phi_i$ depends not only on the coordinates $q$ but also on velocities $\dot{q}$, is  more complicated. There exist many publications devoted to this subject. The discussion of this problem and corresponding references can be found, for example, in \cite{ARNOLD,mann2018lagrangian,JAUME}.

One of the methods  for study mechanical systems with nonholonomic constraint was developed in \cite{KOZLOV}. In this approach one "upgrades" the Lagrangian by adding to it the constraints with corresponding Lagrange multipliers. This method which uses a generalization of Hamilton's principle of stationary action was coined {\em vakonomic mechancs} \cite{KOZLOV,ARNOLD,mann2018lagrangian,JAUME,VAKONOMIC}.
For a system with one constraint $\Phi(q,\dot{q})=0$ the corresponding "upgraded" Lagrangian is
\be
{\CAL L}=L(q,\dot{q})+\chi \Phi(q,\dot{q})\, .
\ee
The variation of the action with respect to Lagrange multiplier $\chi$ reproduces the constraint equation $\Phi=0$, while its  variation over $q$ gives equations containing besides the  coordinates the control function $\chi(t)$. This set of equations determines the constrained motion of the system.

The purpose of this paper is to discuss mechanics with inequality constraints. Such a constraint is a restriction imposed on coordinates $q$ and velocities $\dot{q}$ which are of the form $\Phi(q,\dot{q})\le 0$. We shall demonstrate that  the equations describing the system motion can  be obtained from the following Lagrangian
\be
{\CAL L}=L+\chi (\Phi+\zeta^2)\, ,
\ee
which contains two Lagrange multipliers, $\chi$ and $\zeta$.  Variation of the corresponding action over $\chi$ and $\zeta$ gives
\be
\Phi+\zeta^2=0\hh \chi\zeta=0\, .
\ee
As a result the  motion of the system has two different regimes. If $\Phi\le 0$, the first of these equations defines $\zeta$, while the second equation implies $\chi=0$. When the first equation in saturated and the constraint function $\Phi$ reaches its critical value, $\Phi=0$, one has $\zeta=0$, and the control function $\chi$ can take non-zero value. A transition between these two regimes occurs at  points where the control function $\chi$ vanishes. A similar approach was discussed in \cite{Bertsekas}. In recent publications \cite{FZ1,FZ2,Frolov:2021afd,Frolov:2022fsl} models with inequality constraints imposed on the curvature invariants and their application to the  problem of black hole and cosmological singularities
were discussed.

The paper is organized as follows. In section~II we consider mechanics of a system with one degree of freedom and with an inequality constraint. In section~III we discuss transitions between different regimes and found the corresponding conditions. In section~IV we apply the developed tools for study of a harmonic oscillator with a inequality restriction imposed on its velocity. In section~V we compare a motion of the oscillator in the limiting velocity model with a motion of a relativistic oscillator and demonstrate that there exists similarity between these two models. A case when the constraint function $\Phi$ contains higher than  first derivatives of coordinates is briefly discussed in section~VI. The last section summarizes the obtained results.

\section{Lagrangian mechanics with an inequality constraint}

Our starting point is the following action
\be\n{a1}
S=\int dt {\CAL L},
\ee
where
\be \n{a2}
 {\CAL L}(q,\dot{q},\xi,\zeta)=L(q,\dot{q})+\chi (\Phi(q,\dot{q})+\zeta^2)\, .
\ee
Here and later  we denote by a dot a time derivative.
This action besides the dynamical variable $q(t)$ contains a constraint function $\Phi(q,\dot{q})$ and two Lagrange multipliers $\chi(t)$ and $\zeta(t)$. The variation of $S$ with respect to $\chi(t)$ and $\zeta(t)$ gives
\be \n{a3}
\Phi(q,\dot{q})+\zeta^2=0\hh \chi \zeta=0\, .
\ee
These equations imply that the evolution of the system can have two different regimes:
\begin{itemize}
\item Subcritical phase, where the constraint function $\Phi(q,\dot{q})$ is negative. In this regime the first equation in (\ref{a3}) defines the Lagrange multiplier  $\zeta(t)$, while the second relation implies that $\chi(t)=0$. This means, that during the subcritical phase the system follows the standard Euler-Lagrange equation.
\item Supercritical regime, where $\zeta=0$. At this phase  the constraint equation
\be\n{a4}
\Phi(q,\dot{q})=0
\ee
 is satisfied and the control parameter $\chi(t)$ becomes dynamical.
\end{itemize}

If the system described by action (\ref{a1}) starts its motion in the subcritical regime, then its motion is described by the Euler-Lagrange equation
\be\n{a5}
{\delta L\over \delta q}=0\, .
\ee
Here and later we use the following standard definition of a Lagrange derivative of a function $L(q,\dot{q})$
\be
{\delta L\over \delta q}\equiv {d\over dt}\left( {\partial L\over \partial \dot{q}} \right)-{\partial L\over \partial q}\, .
\ee
Let $q(t)$ be a solution of this equation, then
the constraint function $\Phi(t)=\Phi(q(t),\dot{q}(t))$ calculated on the trajectory is negative. If at some time $t_0$ the constraint function $\Phi(t)$ vanishes the supercritical regimes starts.

At this stage the constraint $\Phi=0$ is valid and the variation of action (\ref{a1})  over $q$ gives
\be \n{a6}
 \dot{\chi}\pi +\chi {\delta \Phi\over \delta q}+{\delta L\over \delta q}= 0\, .
\ee
Here we  denote
\ba\n{a7}
&&\pi\equiv {\partial \Phi\over \partial \dot{q}} \hh {\delta \Phi\over \delta q}\equiv {d\over dt}\left( {\partial \Phi\over \partial \dot{q}} \right)-{\partial \Phi\over \partial q}\, .
\ea
A motion at this phase is specified by two functions of time, $q(t)$ and $\chi(t)$. Function $q(t)$ can be found by solving the constraint equation (\ref{a4}), while equation (\ref{a6}) determines the evolution of the control function $\chi(t)$. This is a first order ordinary differential equation for $\chi(t)$ and its solution is determined if one specifies a value of the control function $\chi$ at some moment of time.

If   the constraint function $\Phi$ reaches its critical value $0$ at some time $t=t_0$, then the $\chi(t_0)=0$. We assume that $\pi(t_0)\ne 0$ then equation (\ref{a6}) with this initial condition has a unique solution. The system can return to its subcritical regime if the control function $\chi(t)$ vanishes again at some later time $t_{1}>t_{0}$.

\section{Condition of the regime change}

Let us discuss a condition when the supercritical solution can return to the subcritical one in more detail.
For this purpose we introduce the following notations
\be  \n{a8}
p= {\partial L\over \partial \dot{q}}\hhh E=p\dot{q}-L\hhh\epsilon=\pi \dot{q}-\Phi\, .
\ee
Let us demonstrate that in the supercritical regime the quantity
\be
{\CAL E}=E+\chi \epsilon\, ,
\ee
is conserved.

It is easy to check that the following relations are valid
\be  \n{a10}
\dot{E}={\delta L\over \delta q}\dot{q}\hh
\dot{\epsilon}={\delta \Phi\over \delta q}\dot{q}\, .
\ee
Using the definition of ${\CAL E}$ and relations (\ref{a10}) one has
\ba \n{a11}
&&\dot{\CAL E}=\dot{E}+\chi \dot{\epsilon}+\dot{\chi}\epsilon \\
&&=\left({\delta L\over \delta q}+\chi {\delta \Phi\over \delta q}+\dot{\chi}\pi\right) \dot{q}-\dot{\chi}\Phi\,  .\\
\ea
Using relation (\ref{a6}) and the constraint equation $\Phi=0$ one gets $\dot{\CAL E}=0$.

The obtained equality means that  ${\CAL E}$ is a conserved  quantity for supercritical solution. By its construction it has the  meaning of the energy of the enlarged system described by the extended action (\ref{a1}) -(\ref{a2}). It differs from the energy $E$ of the unconstrained system by the quantity $\chi \epsilon$. The latter can be interpreted as the contribution of the control field $\chi$  to the total energy of the system. In the subcritical regime  $\chi=0$ and $E=E_0=$const, where $E$ is the usual energy of the unconstrained system. At the transition point $t=t_0$, where $\chi=0$ one has ${\CAL E}=E_0$.

\begin{figure}[!hbt]
    \centering
     \includegraphics[width=0.5 \textwidth]{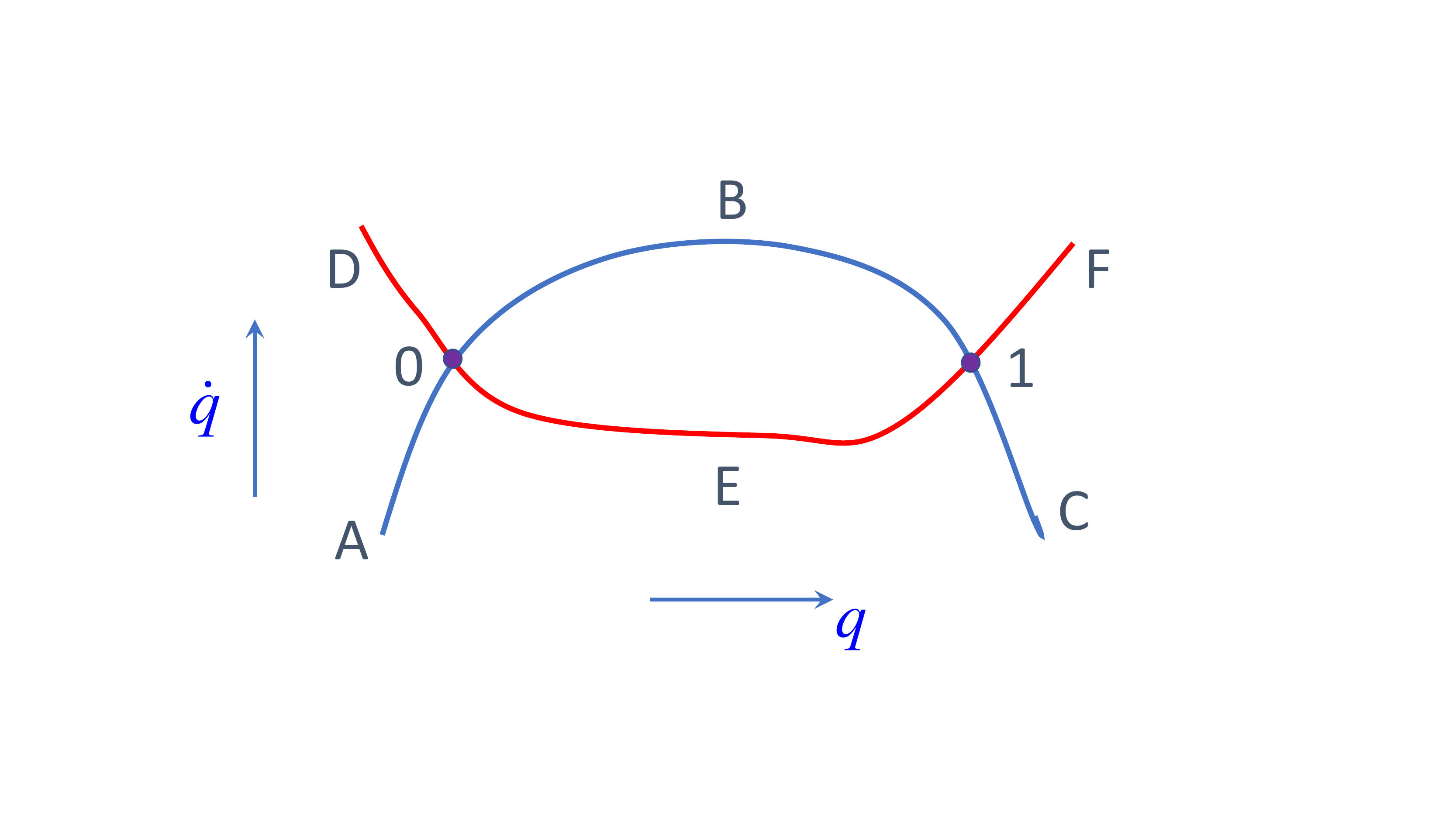}
      \vspace{-50pt}
    \caption{Curves of constant energy, $E=$const, and constraint $\Phi=0$ on the $(q,\dot{q})$ phase plane. }
    \label{F1}
\end{figure}

During the supercritical phase the quantity $E$ depends on time, $E= E(t)$.
It is easy to see that if there exists a later moment of time $t=t_{1}$ where the control parameter $\chi$ becomes zero again, the following condition should be valid
\be \n{a12}
E(t_{1})= E_0\, .
\ee
In other words, the supercritical solution can return to the subcritical regime at the point where the total energy ${\CAL E}$ becomes equal to the energy $E_0$ of the system in the initial subcritical phase.

Fig.~\ref{F1} illustrates a motion of a system with the inequality constraint. It shows curves of constant energy $E$ and of the constraint $\Phi=0$ on the $(q,\dot{q})$ phase plane. Let $ABC$ be $E=$const curve ($E$-curve) and $DEF$ be a constraint curve $\Phi=0$. Point $0$ and $1$ where these curves intersect are transition points between the sub- and supercritical regimes. Suppose that $\Phi$ is positive above the curve $DEF$. Then the system moving from $A$ after reaching the transition point $0$ moves along supercritical trajectory $0E1$. After the second transition point $1$ it continues its motion along a subcritical trajectory $1C$. Since the velocity $\dot{q}$ on $0E1$ is less than the velocity on $0B1$ for the same value of $q$, the time of "travel" between $0$ and $1$
\be
t_{01}=\int_{q_0}^{q_1}{dq\over \dot{q}}
\ee
along the constraint curve is longer than the corresponding time of motion along the $E$-curve.
In the opposite case, where $\Phi$ is positive below the curve $DEF$, the corresponding trajectory is $D0B1F$.

\section{Harmonic oscillator with the limiting velocity constraint}

In order to illustrate the results described in the previous section let us consider a model of a harmonic oscillator with the limiting velocity constraint. Namely, we discuss a system with the following action
\ba
\tilde{S}&=&{1\over 2}\int d\tau \left[  (dx/ d\tau)^2 -\omega^2 x^2 \right.\nonumber\\
&+&\left. \chi  \left( (dx/ d\tau)^2  -V^2+\tilde{\zeta}^2 \right) \right]\, .
\ea
Here $x(\tau)$ is a position of the oscillator at time $\tau$ and $\omega$ is its frequency. The constant $V$ is the value of the limiting velocity and a constraint is chosen so that the following inequality
$|dx/d\tau| \le V$ is valid.
It is convenient to introduce dimensional variables
\be
q=\omega x/V\hh t=\omega\tau\hh \zeta=\tilde{\zeta}/V
\ee
in which
\ba
&&\tilde{S}={V^2\over \omega} S\, , \nonumber \\
&& {S}={1\over 2}\int dt \left[\dot{q}^2 -q^2 +\chi  (\dot{q}^2 -1+{\zeta}^2) \right]\, .
\ea
Here the dot means a derivative with respect to the dimensionless time $t$.

The variation of the action $S$ over $q$, $\chi$ and $\zeta$ gives the following set of equations
\ba
&&(\dot{q}\chi)^.=-(\ddot{q}+q)\, ,\n{4.4}\\
&& \dot{q}^2 -1+\zeta^2=0\, ,\\
&& \zeta\chi=0\, .
\ea
In the subcritical regime $\chi=0$,  $\zeta^2=1- \dot{q}^2$, while the equation (\ref{4.4}) reproduces the standard equation of motion of a non-relativistic harmonic oscillator
\be
\ddot{q}+q=0\, .
\ee
In the supercritical regime
\ba
&&\dot{q}^2 =1\, ,\\
&&\dot{q}\dot{\chi}+\ddot{q}\chi=-(\ddot{q}+q)\n{cc}\, .\n{EQQE}
\ea
The transition between the regimes occurs when the velocity $|\dot{q}|$ reaches its limiting value $1$.

We choose a solution in the subcritical regime in the form
\be \n{qq}
q=-q_0\cos(t)\hh q_0>0\, .
\ee
Then a transition to the supercritical regime  occurs when
\be
\dot{q}\equiv q_0\sin(t)=\pm 1\, .
\ee
This relation can be satisfied only if $q_0\ge 1$\, .  This means that for $q_0<1$ a solution always remains in the subcritical regime and the limiting velocity is not reached.

For $q_0>1$ solution (\ref{qq}) reaches the critical velocity $\dot{q}=1$ for the first time at the moment $t=t_0$
\be
t_0=\arcsin(1/q_0)\, .
\ee
A position of the oscillator at this moment is
\be
q(t_0)=-q_c\hh q_c=\sqrt{q_0^2-1}\, .
\ee
After this time the oscillator follows the supercritical trajectory
\be\n{4.15}
q(t)=-q_c +(t-t_0)\, .
\ee

Equation (\ref{cc}) which determines the evolution of the control function takes the form
\be\n{cq}
\dot{\chi}=-q\, .
\ee
Using (\ref{4.15}) and solving  equation (\ref{cq}) with the initial condition
$\chi(t_0)=0$  one finds
\be
\chi=-{1\over 2}(t-t_0)^2+q_c( t-t_0)\, .
\ee

The control function vanishes again at time $t=t_1>t_0$, where
\be
t_1=t_0 +2q_c\, .
\ee
After this  the solution returns to its subcritical regime where $\chi=0$ and
\be
q=q_0\cos(2(q_c+t_0)-t)\, .
\ee
This subcritical motion continues until the moment of time $t=t_2>t_1$ when the velocity  reaches the critical value $\dot{q}=-1$. For $t_2<t<t_3$ the oscillator moves with a constant critical velocity $\dot{q}=-1$ until a new transition to the subcritical regime occurs at $t=t_3$.

\begin{figure}[!hbt]
    \centering
    \vspace{10pt}
    \includegraphics[width=0.3 \textwidth]{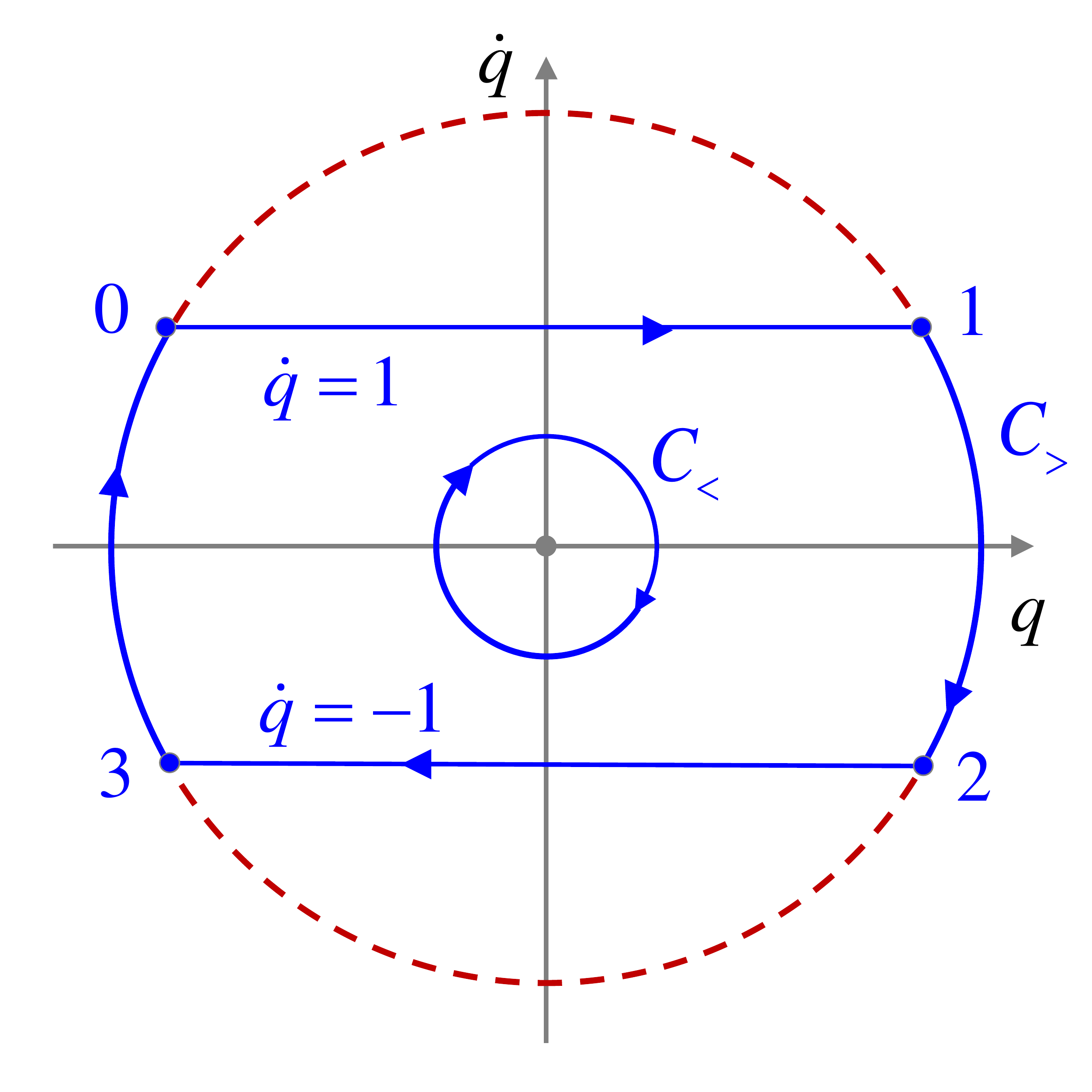}
    \caption{Phase diagram on $(q,\dot{q})$-plane  representing the motion of an oscillator in the limiting velocity model. These plots are shown for two values of the amplitude $q_0=0.5$ (line $C_<$) and $q_0=2.0$ (line $C_>$). }
    \label{F2}
\end{figure}

The phase diagram for the harmonic oscillator in the limiting velocity model is shown at  Fig.~\ref{F2}.
The inner circle $C_<$ describes the motion of the oscillator with $q_0<1$. For this case the supercritical regime is absent. The outer orbit $C_>$ describes the motion of the oscillator with $q_0>1$.
The points 0, 1, 2 and 3 represent  the states where the transitions between subcritical and supercritical  phases occur.

\begin{figure}[!hbt]
    \centering
    \vspace{10pt}
    \includegraphics[width=0.25 \textwidth]{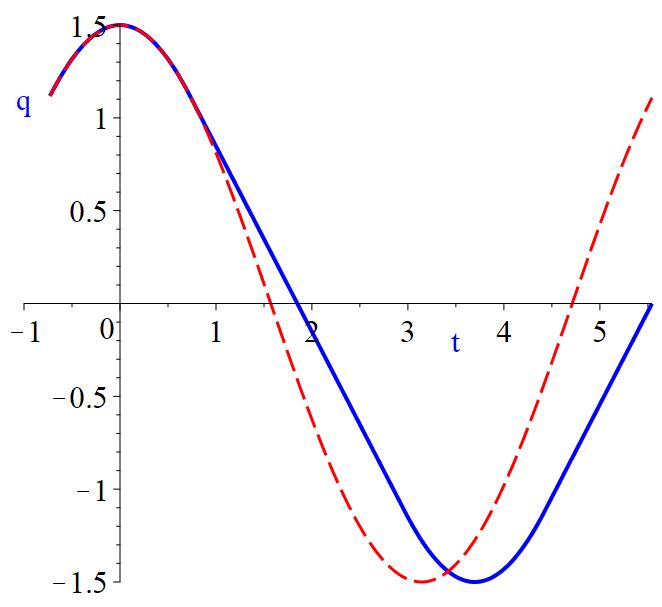}
    \caption{Time dependence of a position of the oscillator with amplitude $q_0=1.5$ in the limiting velocity model is shown by a solid line. The dash line shows a $q(t)$ for an unconstrained oscillator with the same amplitude $q_0$.  }
    \label{F3}
\end{figure}

Fig.\ref{F3} shows the plot of the function $q(t)$ for the oscillator with the velocity constraint (a solid line) and for the standard harmonic oscillator with the same amplitude $q_0>1$ (a dash line).

The motion of the oscillator is periodic. For $q_0<1$ the period is constant and equal to $2\pi$.  For $q_0\ge 1$ the period is
\be
T=4\left[\arcsin(1/q_0)+\sqrt{q_0^2-1}\right]\, .
\ee
Let us note that the period $T$ is always larger than $2\pi$. For $q_0\gg 1$,  $T\sim 4q_0$.

\begin{figure}[!hbt]
    \centering
    \vspace{10pt}
    \includegraphics[width=0.35 \textwidth]{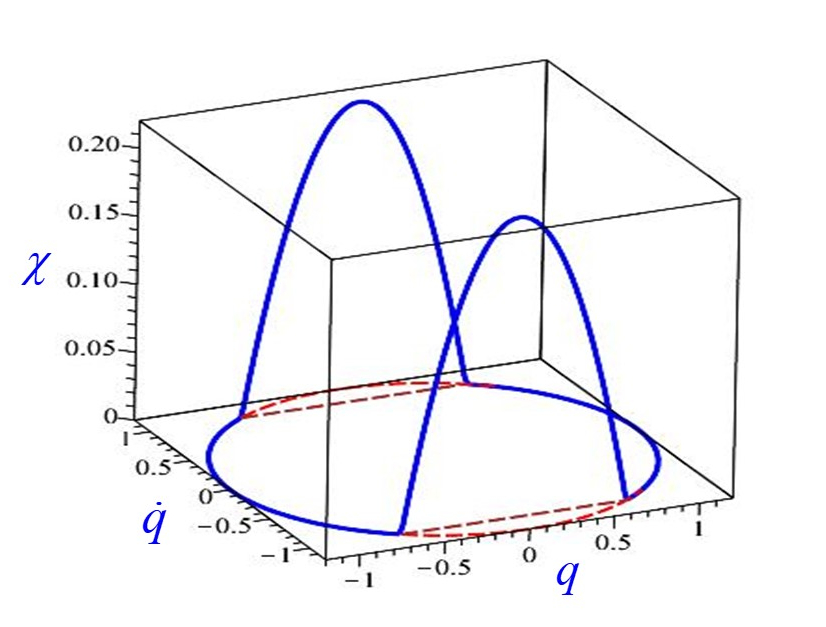}
    \caption{Three-dimensional phase diagram representing the motion of an oscillator in the limiting velocity model. The corresponding 3D coordinates are $(q,\dot{q},\chi)$. This plot is constructed for $q_0=1.2$.}
    \label{F4}
\end{figure}

It is instructive to represent a motion of an oscillator in the limiting velocity model by using 3D phase space with coordinates $(q,\dot{q},\chi)$. Such a phase trajectory is shown at Fig.~\ref{F4}. For the subcritical motion of the system the control function $\chi(t)$ vanishes, and its phase trajectory lies on the $(q,\dot{q})$ plane.
At this stage the energy of the unconstrained harmonic oscillator
\be
E={1\over 2}(\dot{q}^2+q^2)
\ee
is constant. The decrease of the potential energy is compensated by growing of the kinetic energy.

After the system enters the supercritical regime the following quantity is conserved
\be \n{EEE}
{\CAL E}=E+\chi \dot{q}^2\, .
\ee
When the velocity of the system reaches its limiting value, the kinetic energy of the system becomes "frozen", while its potential energy decreases . As a result, the energy $E$ decreases. However, at this stage the control field $\chi$ takes non-zero value and it contributes to the total energy ${\CAL E}$ of the system. This contribution exactly compensates the decrease of $E$. As a result the total energy ${\CAL E}$ during the supercritical phase remains constant. In such a description the field $\chi$ plays the role of some "hidden variable". One can say that this hidden variable at first absorbs a part of the energy of the oscillator and later  releases it. When the total absorbed energy is returned to the oscillator, the field $\chi$ vanishes and total energy ${\CAL E}$ coincides with the standard energy of the unconstrained oscillator $E$. It happens when $\chi$ vanishes and a transition to the subcritical regime occurs.

\section{Limiting velocity oscillator vs relativistic oscillator}

In the previous section we described a motion of the harmonic oscillator in the model with the limiting velocity. In this section we compare this model with the case of a relativistic oscillator where the velocity of the system is naturally restricted by a universal constant $c$ (speed of light). Using the same dimensional units as earlier and putting $c=1$
we write the action for such a relativistic oscillator in the form
\be
S=\int dt L\hh
L=-[\sqrt{1-\dot{q}^2}+{1\over 2}q^2]\, .
\ee
For small velocities, $|\dot{q}|\ll 1$ the Lagrangian $L$ reduces to
\be
L_{nonrel}={1\over 2} [\dot{q}^2-q^2]\, .
\ee
In this relation we omit a constant term which does not affect the equation of motion.
The corresponding Euler-Lagrange equation for the action $S$ is
\be
\dot{p}+q=0\hh p={dL\over d\dot{q}}={\dot{q}\over \sqrt{1-\dot{q}^2}}\, .
\ee
The conserved energy of the relativistic oscillator is
\be\n{EEE}
E=p \dot{q}-L={1\over \sqrt{1-\dot{q}^2}}+{1\over 2}q^2\, .
\ee
The first term in this expression, $(1-\dot{q}^2)^{-1/2}$, is nothing but the Lorentz factor $\gamma$ for a relativistic particle, which for a unit mass coincides with a total (including its rest mass) kinetic energy of the system, while the second term, $q^2$, is the potential energy.

\begin{figure}[!hbt]
    \centering
      \includegraphics[width=0.3 \textwidth]{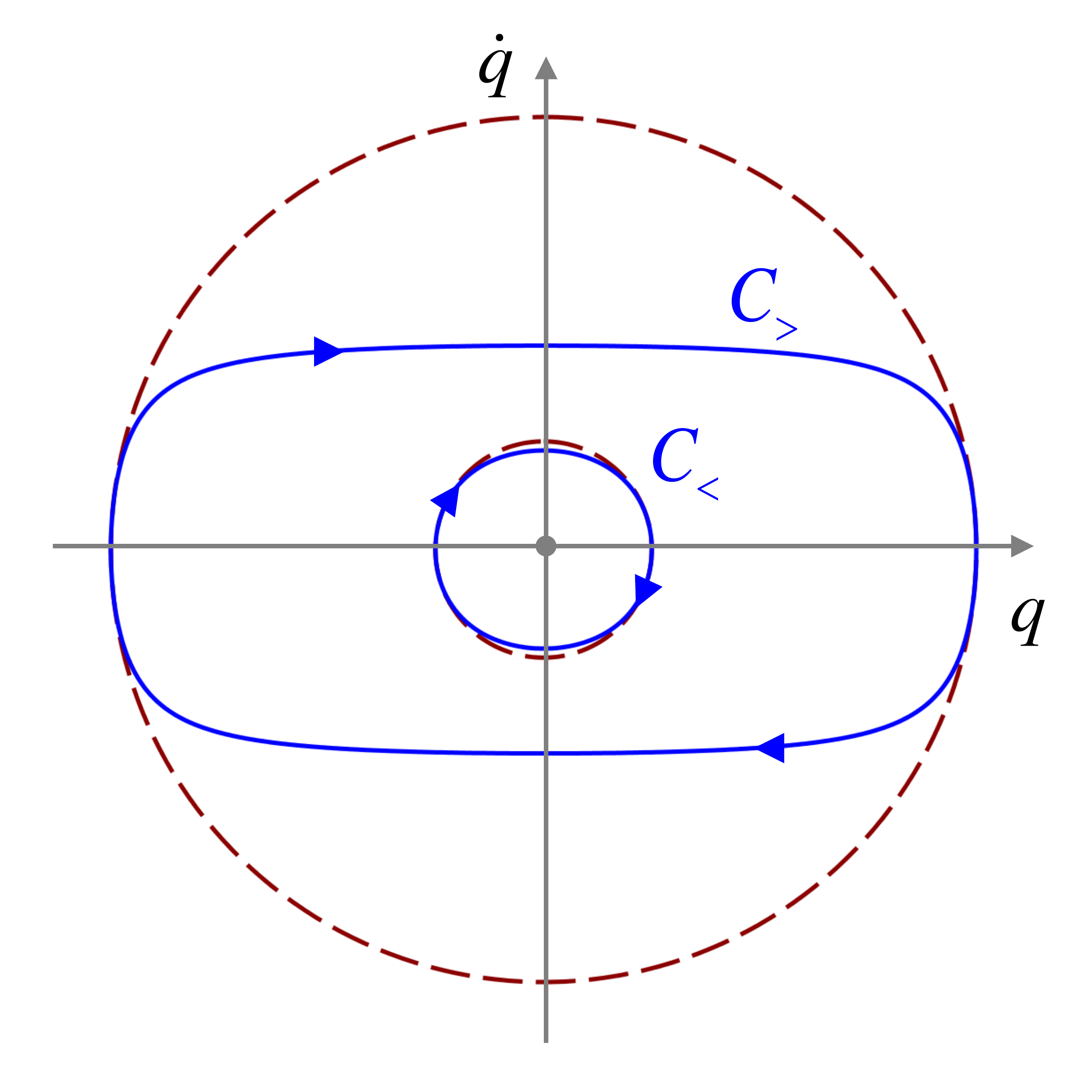}
    \caption{Phase diagram representing the motion of an relativistic oscillator on the $(q,\dot{q})$-plane.  Two plots are shown for two values of the amplitude $q_0=0.5$ (line $C_<$) and $q_0=2.0$ (line $C_>$). For $q_0=0.5$ relativistic effects are small and the phase trajectory $C_<$ of the oscillator  (shown by solid  line) is close to the cycle of radius $0.5$ (shown by a dash line). For the amplitude $q_0$ bigger than 1 the relativistic effects are important. As a result the phase trajectory is squashed in $\dot{q}$ direction. Such a squashed trajectory for  $q_0=2$ is shown  by a solid line $C_>$ .}
    \label{F5}
\end{figure}

The motion is bounded. Let us denote the maximal value of the coordinate $q$ for a given energy $E$ by $q_0$, then one has
\be
E=1+{1\over 2}q_0^2\, ,
\ee
and the equation (\ref{EEE}) takes the form
\be
\gamma\equiv (1-\dot{q}^2)^{-1/2}=1+z\hh z={1\over 2}(q_0^2-q^2)\, .
\ee
Solving this equation for $\dot{q}$ one gets
\be \n{rel_q}
\dot{q}=\pm {\sqrt{z(2+z)}\over 1+z}\, .
\ee
This relation allows one to plot a  phase diagram for the motion of the relativistic oscillator in $(q,\dot{q})$ plane.
 This diagram is shown at Fig.~\ref{F5}.

The  velocity of the oscillator has a maximal value at $q=0$
\be
\dot{q}_0=\pm {\sqrt{z_0(2+z_0)}\over 1+z_0}\hh z_0={1\over 2}q_0^2\, .
\ee
It is easy to check that $|\dot{q}_0|<1$ and it becomes close to $\pm 1$ for large values of $q_0$.

\begin{figure}[!hbt]
    \centering
      \includegraphics[width=0.23 \textwidth]{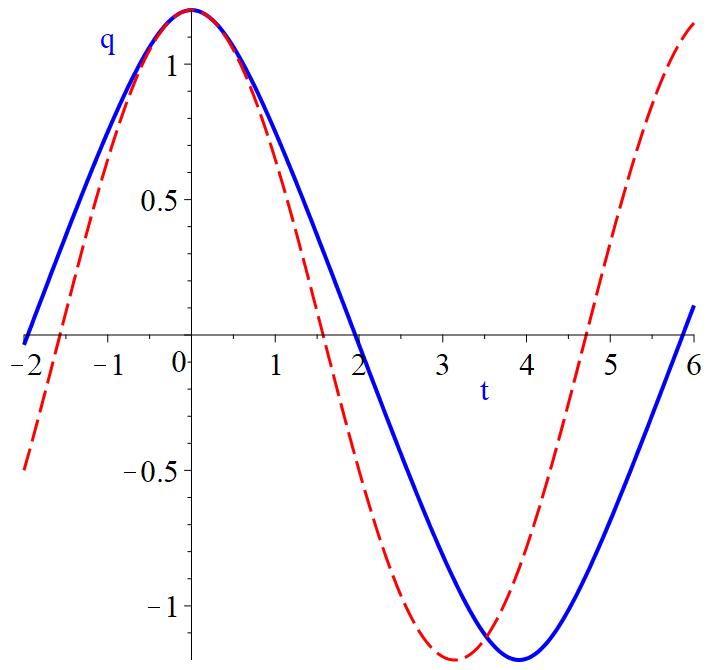}
    \caption{Time dependence of a position of the relativistic oscillator with amplitude $q_0=1.2$ is shown by a solid line. The dash line shows a position of a non-relativistic oscillator with the same amplitude. }
    \label{F6}
\end{figure}

By integrating the first order differential equation (\ref{rel_q}) one can find a position of the oscillator at time $t$, $q(t)$. To illustrate its behavior we show at Fig.~\ref{F6}
function $q(t)$ for $q_0=1.2$. For comparison, this figure also shows the function $q(t)$ for a similar non-relativistic oscillator with the same amplitude $q_0$ (a dash line).

The motion of the relativistic oscillator is periodic. The corresponding period is
\be
T_{rel}=2\sqrt{2}\int_0^{z_0} {dz (1+z)\over \sqrt{z(2+z)(z_0-z)}}\, .
\ee
This integral can be calculated analytically with the following result
\be
T_{rel}={4\sqrt{2}\over \sqrt{z_0+2}}
\left[(z_0+2)E\left(\sqrt{z_0\over z_0+2}\right)-
K\left(\sqrt{z_0\over z_0+2}\right)\right]\, .
\ee
Here $E(x)$ and $K(x)$ are complete elliptic integrals. For $q_0\gg 1$
\be
T_{rel}\sim 4q_0\, .
\ee
This result has a simple explanation: For a large amplitude $q_0$ the system very fast reaches the velocity close to the speed of light, which in our units is $1$.

\begin{figure}[!hbt]
    \centering
      \includegraphics[width=0.3 \textwidth]{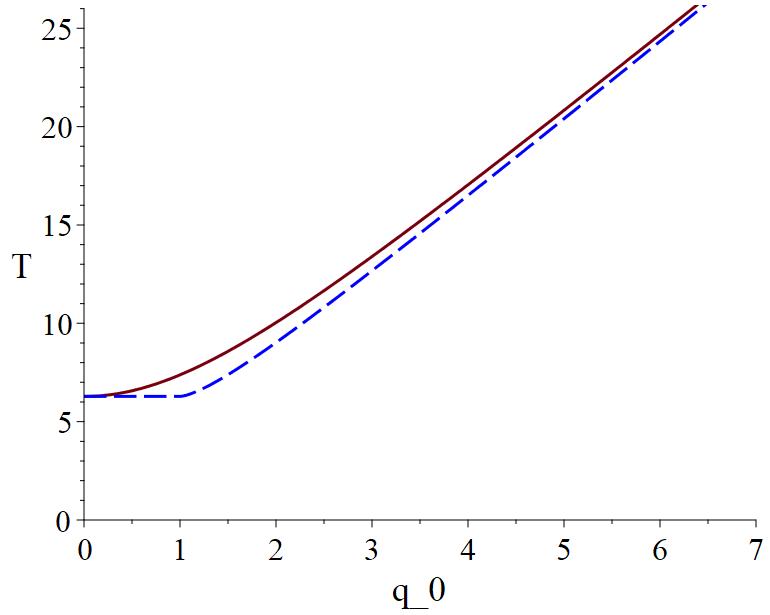}
    \caption{Periods of the  relativistic oscillator (solid line) and the oscillator in the limiting velocity model (dash line) as functions of their amplitude $q_0$.}
    \label{F7}
\end{figure}

Fig.~\ref{F7} shows the dependence of the period $T$ of the relativistic oscillator as a function of its amplitude $q_0$ (solid line). A dash line at this plot shows the dependence of the period of the oscillator in the limiting velocity model on its amplitude $q_0$. One can see that for both cases the corresponding lines are quite similar.

\section{Inequality constraints with higher than first derivatives}

Till now we considered a case when the constraint function $\Phi$ depends only on $q$ and $\dot{q}$. Let us discuss now what happens if the constraint function contains  second or higher derivatives of $q$. Let us assume first that $\Phi$ contains second derivative of $q$, which enters linearly. Then the Lagrangian ${\CAL L}$ which enters the action (\ref{a1}) is
\be
{\CAL L}=L(q,\dot{q})+\chi (\Phi+\zeta^2)\hh
\Phi=\ddot{q}-\varphi(q,\dot{q})\, .
\ee
Such a constraint means that the acceleration $\ddot{q}$ is restricted from above by the quantity $\varphi(q,\dot{q})$. If a subcritical solution reaches a point where $\ddot{q}=\varphi(q,\dot{q})$ it enters the supercritical regime where the following equations are valid
\ba
&& \ddot{q}=\varphi(q,\dot{q})\, ,\n{ddd}\\
&&{\delta {\CAL L}\over \delta q}\equiv {\delta {L}\over \delta q}-{\delta {(\chi\varphi )}\over \delta q}-\ddot{\chi}=0\, .\n{LAG}
\ea
A trajectory $q(t)$ can be found by solving the first of these equations.
The second equation defines the evolution of the control function $\chi$. It can be written in the form
\be\n{sode}
\ddot{\chi}+\varphi_{,\dot{q}}\dot{\chi}+  {\delta \varphi\over \delta q}\chi={\delta { L}\over \delta q}\, .
\ee
This is a second order linear inhomogeneous ordinary differential equation. Its solutions are uniquely determined by specifying the initial values of  $\chi$ and $\dot{\chi}$ at some moment of time. In the subcritical regime
the control function $\chi$ identically vanish. Equation (\ref{sode}) implies that
if $\varphi_{,\dot{q}}$ and $\delta\varphi/\delta q$ are finite, then
neither $\chi$ no $\dot{\chi}$ can jump. Hence, at the transition point point one has
\be \n{6.5}
\chi=\dot{\chi}=0\, .
\ee
 After substitution of a solution $q(t)$  of the constraint equation into the right-hand side of (\ref{sode}) and
using this initial conditions one can find the time-dependence of the control function in the supercritical regime.

One can check that during the supercritical phase the following quantity remains constant
\be \n{EEE}
{\CAL E}=E-\dot{q}\dot{\chi}+[\varphi-\varphi_{,\dot{q}}\dot{q}]\chi \hh
E=\dot{q}L_{,\dot{q}}-L \, .
\ee
To prove that $\dot{\CAL E}=0$ it is sufficient to use  equation (\ref{ddd}) and the following relation
\be
\dot{E}={\delta L\over \delta q} \dot{q}\, .
\ee
As earlier one can consider ${\CAL E}$ as an "upgraded" version of the energy, which besides the energy $E$ of the original unconstrained system contains a contribution of the energy related to the control function $\chi(t)$. Motion along the constraint  $\Phi=0$ results in the change of $E$, which is compensated by the  contribution of the control parameter variable to the total energy.

Let us demonstrate that in a general case, if the solution has entered in the supercritical regime it cannot return again to its subcritical phase. Let us suppose opposite: at some moment of time $t_1$ the supercritical solution leaves the constraint. Since for $t>t_1$ the control function $\chi(t)$ should identically vanish, at the point of the transition one has $\chi=\dot{\chi}=0$. If $\varphi_{,\dot{q}}$ and $\delta\varphi/\delta q$ are finite the equation (\ref{sode}) implies that the functions $\dot{\chi}(t)$ and $\chi(t)$ cannot jump. But for a general solution of (\ref{sode}) which vanishes at $t=t_1$  its first derivative does not necessary vanish at this point and the condition $\chi=\dot{\chi}=0$ is not valid . This property makes the case when the constraint function $\Phi$ contains second derivatives quite different from earlier considered model with $\Phi(q,\dot{q})$.

One can expect that this is a generic property of models with inequality constraints which contain second and/or higher derivatives of $q$.  In other words, if  a solution enters from subcritical regime to the supercritical one, then it is highly likely that the solution remains in the supercritical phase forever.

\section{A harmonic oscillator in a limiting acceleration model}

To illustrate properties of a system with an inequality constrain containing second  derivatives let us consider a simple model of a non-relativistic harmonic oscillator with a limiting acceleration. Namely, we assume that
the acceleration of the oscillator should always be smaller than some positive quantity, which we denote by $a$. We choose the corresponding constraint function $\Phi$ as follows
\be
\Phi=\ddot{q}-a\, ,
\ee
and write the action in the form
\be
{S}=\int dt \left[{1\over 2}(\dot{q}^2 -q^2) +\chi  (\Phi+{\zeta}^2) \right]\, .
\ee

We assume that the oscillator starts its motion at $t=0$ with $q=0$ and its velocity is $\dot{q}=-q_0$
Then the equation of motion of such unconstrained oscillator in the subcritical regime is
\be
q(t)=-q_0\sin(t)\, .
\ee
Its acceleration $\ddot{q}=q_0\sin(t)$ grows in time. If $q_0<a$ the oscillator acceleration always remains less than the critical one, so that its motion remains subcritical forever.

In the case when $q_0>a$ its behavior is quite different.
Denote by $t_0$ the moment of time where
\be
\sin(t_0)=a/q_0\, .
\ee
A coordinate $q$ and  velocity of the oscillator at this moment are
\be
q(t_0)=-a\hh \dot{q}(t_0)=-\sqrt{q_0^2-a^2}, .
\ee
After $t=t_0$ the oscillator moves with a constant acceleration $a$ and one has
\be\n{qqq}
q(t)={1\over 2}a(t-t_0)^2-\sqrt{q_0^2-a^2}(t-t_0)-a\, .
\ee
The equation (\ref{sode}) for the control function $\chi(t)$ takes the form
\be \n{ddqq}
\ddot{\chi}=\ddot{q}+q\, ,
\ee
where $q(t)$ is given by (\ref{qqq}).
The equation (\ref{ddqq})  can be easily integrated and one has
\be
\chi(t)={1\over 6}(t-t_0)^3 \left[{1\over 4}a(t-t_0)-\sqrt{q_0^2-a^2}\right]\, .
\ee
We used here the initial conditions $\chi(t_0)=\dot{\chi}(t_0)=0$.

The control function  becomes zero again at $t=t_1$
\be
t_1-t_0={4\over a}\sqrt{q_0^2-a^2}\, .
\ee
However, at this point the velocity of the oscillator does not vanish and is
\be
\dot{\chi}(t_1)={8\over 3} \left[ (q_0/a)^2-1\right]^{3/2}\ne 0\, .
\ee
Thus the conditions $\chi(t_1)=0$ and $\dot{\chi}(t_1)=0$ cannot be satisfied simultaneously.  This means that in the model with a limiting acceleration the oscillator  after it  enters from the subcritical regime to the supercritical one remains in the supercritical regime forever.

\section{Summary}

In this paper we discussed properties of dynamical systems with inequality constraints. It was demonstrated that such constraints can be taken into account by proper modification of the original action of the unconstrained theory. We first discussed a theory with Lagrangian $L(q,\dot{q})$ and inequality constraint of the form $\Phi(q,\dot{q})\le 0$. The dynamics of such a system is describes by the extended Lagrangian ${\CAL L}=L+\chi (\Phi+\zeta^2)$ which contains two Lagrange multipliers $\chi(t)$ ad $\zeta(t)$. The motion of a system  has two regimes: sub- and supercritical ones. A transition between these regimes occurs at the points of intersection of the constant energy $E$ line and the constraint curve $\Phi$ on the phase plane $(q,\dot{q})$. For the subcritical solution, as well at the
 transition point, the control parameter $\chi(t)$ vanishes. During the supercritical phase the energy $E(q,\dot{q})$  does not conserve, while the total energy ${\CAL E}(q,\dot{q},\chi)$, which contains a contribution of the control function, is an integral of motion.

To illustrate properties of dynamical systems with an inequality constraint we considered a simple model of a harmonic oscillator with an imposed condition that it velocity cannot be larger then some fixed value. Using the dimensionless units this velocity can be chosen equal to  1. We showed that if the amplitude $q_0$ of the oscillator is less than 1, its periodic motion does not have a supercritical phase, while for $q_0>1$ it contains both phases, sub- and supercritical ones. We compared the motion of the oscillator in the limiting velocity model with the motion of a relativistic oscillator and demonstrated that their properties are quite similar. In this sense the model with a limiting velocity constraint can be considered as a "poor person" version of the special relativity. Certainly this model does not pretend to substitute it. It plays a role of some phenomenological model which  captures some  features of the special relativity where the limiting nature of the speed of light is important.

At the end of the paper we discussed inequality constrains which contain higher than the first derivations of the dynamical variables. For such systems the differential equation for the control function is of the second or higher order in derivatives and for the transition between sub- and supercritical regimes more than one condition should be satisfied. The conditions can be satisfied either for the transition from sub- to supercritical phase or for the inverse transition. However, one can expect that  a situation when there exist more than one transitions between the sub- and supercritical phases is rather special and it requires some additional conditions.

In our discussion we focused on systems with one degrees of freedom with a single inequality constraint. It is possible to extend this analysis to more general systems which have more than one degree of freedom and several inequality constrains. One can expect that the behavior of such systems is more complicated. In particular, besides a subcritical regime there may exist several supercritical regimes with possible transitions between them. Study of such systems is a quite interesting problem.

Simple examples discussed in this paper illustrate  some general  features of the models with inequality constraints.
They can be useful in discussion of the more general and physically more interesting cases.
 For example, in the recent papers \cite{FZ1,FZ2,Frolov:2021afd,Frolov:2022fsl} the limiting curvature models were proposed and discussed in connection with the singularity problems in cosmology and in black holes in Einstein gravity. The basic idea of this approach is to find some robust predictions of (still unknown) theory of gravity where curvature cannot be infinitely large. In this sense, the gravity models with the limiting curvature serve as some phenomenological models which can be used for study predictions of a more fundamental modified gravity theory. Models with inequality constraints considered in the paper are rather simple and were used only to illustrate some simple properties of such systems. It would be interesting to apply a similar approach for other physically interesting problems.

\section*{Acknowledgments}

The authors are grateful to  the Natural Sciences and Engineering Research Council of Canada for its
 financial support.
V.F.\  also thanks the Killam Trust for its financial support. The authors thank Dr.Andrei Zelnikov for his help with preparation of some of the figures.
\vfill


\begin{thebibliography}{12}%
\makeatletter
\providecommand \@ifxundefined [1]{%
 \@ifx{#1\undefined}
}%
\providecommand \@ifnum [1]{%
 \ifnum #1\expandafter \@firstoftwo
 \else \expandafter \@secondoftwo
 \fi
}%
\providecommand \@ifx [1]{%
 \ifx #1\expandafter \@firstoftwo
 \else \expandafter \@secondoftwo
 \fi
}%
\providecommand \natexlab [1]{#1}%
\providecommand \enquote  [1]{``#1''}%
\providecommand \bibnamefont  [1]{#1}%
\providecommand \bibfnamefont [1]{#1}%
\providecommand \citenamefont [1]{#1}%
\providecommand \href@noop [0]{\@secondoftwo}%
\providecommand \href [0]{\begingroup \@sanitize@url \@href}%
\providecommand \@href[1]{\@@startlink{#1}\@@href}%
\providecommand \@@href[1]{\endgroup#1\@@endlink}%
\providecommand \@sanitize@url [0]{\catcode `\\12\catcode `\$12\catcode
  `\&12\catcode `\#12\catcode `\^12\catcode `\_12\catcode `\%12\relax}%
\providecommand \@@startlink[1]{}%
\providecommand \@@endlink[0]{}%
\providecommand \url  [0]{\begingroup\@sanitize@url \@url }%
\providecommand \@url [1]{\endgroup\@href {#1}{\urlprefix }}%
\providecommand \urlprefix  [0]{URL }%
\providecommand \Eprint [0]{\href }%
\providecommand \doibase [0]{http://dx.doi.org/}%
\providecommand \selectlanguage [0]{\@gobble}%
\providecommand \bibinfo  [0]{\@secondoftwo}%
\providecommand \bibfield  [0]{\@secondoftwo}%
\providecommand \translation [1]{[#1]}%
\providecommand \BibitemOpen [0]{}%
\providecommand \bibitemStop [0]{}%
\providecommand \bibitemNoStop [0]{.\EOS\space}%
\providecommand \EOS [0]{\spacefactor3000\relax}%
\providecommand \BibitemShut  [1]{\csname bibitem#1\endcsname}%
\let\auto@bib@innerbib\@empty
\bibitem [{\citenamefont {Goldstein}\ \emph {et~al.}(2002)\citenamefont
  {Goldstein}, \citenamefont {Poole},\ and\ \citenamefont
  {Safko}}]{goldstein2002}%
  \BibitemOpen
  \bibfield  {author} {\bibinfo {author} {\bibfnamefont {H.}~\bibnamefont
  {Goldstein}}, \bibinfo {author} {\bibfnamefont {C.}~\bibnamefont {Poole}}, \
  and\ \bibinfo {author} {\bibfnamefont {J.}~\bibnamefont {Safko}},\
  }\href@noop {} {\emph {\bibinfo {title} {{Classical Mechanics}}}}\ (\bibinfo
  {publisher} {Addison Wesley},\ \bibinfo {year} {2002})\BibitemShut {NoStop}%
\bibitem [{\citenamefont {Arnold}\ \emph {et~al.}(1989)\citenamefont {Arnold},
  \citenamefont {Vogtmann},\ and\ \citenamefont {Weinstein}}]{arnold89}%
  \BibitemOpen
  \bibfield  {author} {\bibinfo {author} {\bibfnamefont {V.}~\bibnamefont
  {Arnold}}, \bibinfo {author} {\bibfnamefont {K.}~\bibnamefont {Vogtmann}}, \
  and\ \bibinfo {author} {\bibfnamefont {A.}~\bibnamefont {Weinstein}},\
  }\href@noop {} {\emph {\bibinfo {title} {Mathematical Methods of Classical
  Mechanics}}},\ Graduate texts in mathematics\ (\bibinfo  {publisher}
  {Springer},\ \bibinfo {year} {1989})\BibitemShut {NoStop}%
\bibitem [{\citenamefont {{Arnold}}(1988)}]{ARNOLD}%
  \BibitemOpen
  \bibfield  {author} {\bibinfo {author} {\bibfnamefont {V.~I.}\ \bibnamefont
  {{Arnold}}},\ }\href@noop {} {\emph {\bibinfo {title} {{Dynamical systems
  III.}}}},\ Vol.~\bibinfo {volume} {3}\ (\bibinfo {year} {1988})\BibitemShut
  {NoStop}%
\bibitem [{\citenamefont {{Mann}}(2018)}]{mann2018lagrangian}%
  \BibitemOpen
  \bibfield  {author} {\bibinfo {author} {\bibfnamefont {P.}~\bibnamefont
  {{Mann}}},\ }\href@noop {} {\emph {\bibinfo {title} {Lagrangian and
  Hamiltonian Dynamics}}}\ (\bibinfo  {publisher} {Oxford University Press},\
  \bibinfo {year} {2018})\BibitemShut {NoStop}%
\bibitem [{\citenamefont {{Jaume Llibre and Rafael Ramirez}}(2016)}]{JAUME}%
  \BibitemOpen
  \bibfield  {author} {\bibinfo {author} {\bibnamefont {{Jaume Llibre and
  Rafael Ramirez}}},\ }\href@noop {} {\emph {\bibinfo {title} {Inverse problems
  in ordinary differential equations and applications}}},\ \bibinfo {series}
  {Progress in Mathematics}, Vol.\ \bibinfo {volume} {313}\ (\bibinfo
  {publisher} {Birkhauser},\ \bibinfo {year} {2016})\BibitemShut {NoStop}%
\bibitem [{\citenamefont {{Kozlov}}(1983)}]{KOZLOV}%
  \BibitemOpen
  \bibfield  {author} {\bibinfo {author} {\bibfnamefont {V.~V.}\ \bibnamefont
  {{Kozlov}}},\ }\href@noop {} {\bibfield  {journal} {\bibinfo  {journal}
  {Soviet Physics. Doklady. Translation of the physics sections of Doklady
  Akademii Nauk SSSR}\ }\textbf {\bibinfo {volume} {28}},\ \bibinfo {pages}
  {594–600} (\bibinfo {year} {1983})}\BibitemShut {NoStop}%
\bibitem [{\citenamefont {Cortes}\ \emph {et~al.}(2000)\citenamefont {Cortes},
  \citenamefont {León}, \citenamefont {Martin~de Diego},\ and\ \citenamefont
  {Martínez}}]{VAKONOMIC}%
  \BibitemOpen
  \bibfield  {author} {\bibinfo {author} {\bibfnamefont {J.}~\bibnamefont
  {Cortes}}, \bibinfo {author} {\bibfnamefont {M.}~\bibnamefont {León}},
  \bibinfo {author} {\bibfnamefont {D.}~\bibnamefont {Martin~de Diego}}, \ and\
  \bibinfo {author} {\bibfnamefont {S.}~\bibnamefont {Martínez}},\ }\href@noop
  {} {\bibfield  {journal} {\bibinfo  {journal} {SIAM Journal on Control and
  Optimization}\ }\textbf {\bibinfo {volume} {41}},\ \bibinfo {pages} {1}
  (\bibinfo {year} {2000})}\BibitemShut {NoStop}%
\bibitem [{\citenamefont {Bertsekas}(1996)}]{Bertsekas}%
  \BibitemOpen
  \bibfield  {author} {\bibinfo {author} {\bibfnamefont {D.~P.}\ \bibnamefont
  {Bertsekas}},\ }\href@noop {} {\emph {\bibinfo {title} {Constrained
  Optimization and Lagrange Multiplier Methods (Optimization and Neural
  Computation Series)}}}\ (\bibinfo  {publisher} {Athena Scientific},\ \bibinfo
  {year} {1996})\BibitemShut {NoStop}%
\bibitem [{\citenamefont {Frolov}\ and\ \citenamefont
  {Zelnikov}(2021{\natexlab{a}})}]{FZ1}%
  \BibitemOpen
  \bibfield  {author} {\bibinfo {author} {\bibfnamefont {V.~P.}\ \bibnamefont
  {Frolov}}\ and\ \bibinfo {author} {\bibfnamefont {A.}~\bibnamefont
  {Zelnikov}},\ }\href@noop {} {\bibfield  {journal} {\bibinfo  {journal}
  {Journal of High Energy Physics}\ }\textbf {\bibinfo {volume} {2021}},\
  \bibinfo {pages} {154} (\bibinfo {year} {2021}{\natexlab{a}})}\BibitemShut
  {NoStop}%
\bibitem [{\citenamefont {Frolov}\ and\ \citenamefont
  {Zelnikov}(2021{\natexlab{b}})}]{FZ2}%
  \BibitemOpen
  \bibfield  {author} {\bibinfo {author} {\bibfnamefont {V.~P.}\ \bibnamefont
  {Frolov}}\ and\ \bibinfo {author} {\bibfnamefont {A.}~\bibnamefont
  {Zelnikov}},\ }\href@noop {} {\bibfield  {journal} {\bibinfo  {journal}
  {Phys. Rev. D}\ }\textbf {\bibinfo {volume} {104}},\ \bibinfo {pages}
  {104060} (\bibinfo {year} {2021}{\natexlab{b}})}\BibitemShut {NoStop}%
\bibitem [{\citenamefont {Frolov}\ and\ \citenamefont
  {Zelnikov}(2022)}]{Frolov:2021afd}%
  \BibitemOpen
  \bibfield  {author} {\bibinfo {author} {\bibfnamefont {V.~P.}\ \bibnamefont
  {Frolov}}\ and\ \bibinfo {author} {\bibfnamefont {A.}~\bibnamefont
  {Zelnikov}},\ }\href@noop {} {\bibfield  {journal} {\bibinfo  {journal}
  {Phys. Rev. D}\ }\textbf {\bibinfo {volume} {105}},\ \bibinfo {pages}
  {024041} (\bibinfo {year} {2022})}\BibitemShut {NoStop}%
\bibitem [{\citenamefont {Frolov}(2022)}]{Frolov:2022fsl}%
  \BibitemOpen
  \bibfield  {author} {\bibinfo {author} {\bibfnamefont {V.~P.}\ \bibnamefont
  {Frolov}},\ }\href@noop {} {\bibfield  {journal} {\bibinfo  {journal} {Int.
  J. Mod. Phys. A}\ }\textbf {\bibinfo {volume} {37}},\ \bibinfo {pages}
  {2243009} (\bibinfo {year} {2022})}\BibitemShut {NoStop}%
\end{thebibliography}

%

\end{document}